\documentclass[sigconf, nonacm]{acmart}              
\usepackage{bm}

\AtBeginDocument{%
  \providecommand\BibTeX{{%
    \normalfont B\kern-0.5em{\scshape i\kern-0.25em b}\kern-0.8em\TeX}}}

\acmConference[BeyondFacts'24]{4th International Workshop on Computational Methods for Online Discourse Analysis}{May 13--17, 2024}{Singapore}

\setcopyright{acmcopyright}
\copyrightyear{2025}
\acmYear{2025}
\acmDOI{XXXXXXX.XXXXXXX}
\acmPrice{00.00}
\acmISBN{978-1-4503-9419-2/24/05}

\begin{document}
\title{Who shapes Web standards? Uncovering the main topics of interest in the W3C}


\author{Henrique S. Xavier}
\email{hxavier@nic.br}
\orcid{0000-0002-9601-601X}
\affiliation{%
  \institution{NIC.br}
  \streetaddress{Av. das Nações Unidas, 11541, 7º andar}
  \city{São Paulo}
  \state{SP}
  \country{Brazil}
  \postcode{04578-000}
}

\author{Beatriz Rocha}
\email{biarocha@nic.br}
\affiliation{
  \institution{NIC.br}
  \streetaddress{Av. das Nações Unidas, 11541, 7º andar}
  \city{São Paulo}
  \state{SP}
  \country{Brazil}
  \postcode{04578-000}
}

\author{Diogo Cortiz}
\email{diogo@nic.br}
\affiliation{%
  \institution{NIC.br}
  \streetaddress{Av. das Nações Unidas, 11541, 7º andar}
  \city{São Paulo}
  \state{SP}
  \country{Brazil}
  \postcode{04578-000}
}

\renewcommand{\shortauthors}{Xavier et al.}

\begin{abstract}
This paper identifies the primary topics of interest of organizations participating in the World Wide Web Consortium (W3C), the leading standards body for the Web. Using publicly available data from the W3C website, we analyze the participation of member organizations in W3C groups, treating the number of representatives allocated to each group as a proxy for their interests. By applying topic modeling and similarity analysis to these participation patterns, we uncover clusters of related groups and shared priorities among organizations. The results reveal five prominent areas of focus -- Web, Ads \& Privacy; High Performance; Credentials \& Web of Things; Accessibility; and Payments -- and show that large enterprises, particularly those based in the United States, dominate participation in core Web development and advertising-related topics, while Japanese organizations are more active in the Web of Things. These findings offer insights into how various stakeholders influence the standardization process and how the Web may evolve in the coming years.   
\end{abstract}



\received{05 November 2025}

\maketitle

\section{Introduction}
\label{sec:intro}

The Web has profoundly transformed how we communicate, study, work, and live, becoming indispensable to countless human activities. Since its creation in 1989, it has evolved dramatically, often in unexpected directions. Originally designed as a decentralized academic information system for sharing documents \cite{BernersLee1990}, the Web has expanded to host interoperable, dynamic, interactive, and real-time applications, including Software-as-a-Service (SaaS), e-commerce, social media, and streaming platforms. It has also enabled new practices such as user monitoring, testing, and nudging \cite{Zuboff2019}, as well as profiting from crowdsourcing, collective intelligence, user-generated content, data mining, and targeted advertising \cite{OReilly2005}. Moreover, user activity and content have become increasingly concentrated within a few large for-profit platforms \cite{Xavier2024}. 

From the history of the Web, it is evident that anticipating its evolution holds significant economic and social value. Although challenging, examining the Web's current state -- its key actors and their focus areas -- can reveal clues about its possible future directions. Several studies have pursued this line of inquiry using different indicators. Xavier analyzed online traffic data from SimilarWeb, identifying major industries, pervasive practices, and the main controllers of popular websites \cite{Xavier2024}, while Graux \& Orlandi examined the topics addressed in the Web Conference series to trace historical and emerging trends in Web-related research \cite{Graux2022}. These studies highlight the prominent role of major U.S. technology companies in shaping research and user experience on the Web, as well as the importance of video streaming services, business models based on user- or third-party-generated content, and the widespread use of artificial intelligence -- particularly recommender systems. In contrast, the Semantic Web, once envisioned as the Web's future \cite{BernersLee2001}, has largely stagnated in research and seen limited practical adoption \cite{Hogan2020}.

We aim to extend these analyses by examining the participation of different actors in the World Wide Web Consortium (W3C), the primary standards development organization (SDO) for the Web. The W3C develops many standards and best practices that shape the Web, with members ranging from large corporations to small businesses, non-profit organizations, government agencies, and academic institutions. By studying the activities of these actors within the W3C, we seek to uncover their interests and priorities, as well as the potential implications of their work for the Web's future.

The main W3C activities occur within groups\footnote{\url{https://www.w3.org/groups}}, which are classified as either permanent or temporary. There are four permanent groups with governance and oversight roles: the Advisory Board (AB), the Technical Architecture Group (TAG), the Advisory Committee (AC), and the Board of Directors (BoD). Temporary groups, in turn, are established to address specific topics related to Web standards and practices and are expected to disband once their objectives are met. These are divided into chartered and non-chartered groups. Participation in chartered groups is restricted to W3C members, staff, and invited experts, while non-chartered groups are open to anyone. Chartered groups include Working Groups (WG) and Interest Groups (IG), whereas non-chartered groups comprise Community Groups (CG) and Business Groups (BG). Although only WGs are authorized to produce W3C Recommendations (i.e., standards)\footnote{\url{https://www.w3.org/standards/types/\#x5-standard}}, all group types play important roles in the Web's development. Community and Business groups, for example, serve as incubators for new ideas and technologies, often generating momentum for later standardization efforts\footnote{\url{https://www.w3.org/community/about/faq}} \cite{Harcourt2020}.

Previous studies on the W3C have shown that, although it includes members from academia, government, and civil society, large corporations remain the dominant actors, followed by smaller companies \cite{Gamalielsson2016, Halpin2017, Harcourt2020, Mason2019}. This pattern is consistent with other SDOs. According to \cite{Harcourt2020}, most standards are submitted to the W3C after being developed and implemented elsewhere, typically when the need for interoperability becomes evident. The authors note that ``most salient work within the W3C is led mainly by browser companies such as Google and Microsoft, manufacturers like Intel, and user companies such as Facebook.'' At the same time, ``Comcast and Netflix are also important actors particularly within specifications for television.'' A 2016 analysis of the affiliations of W3C standards editors found collaborations between universities and large and small companies, as well as the dominance of U.S.-based organizations \cite{Gamalielsson2016}. Gupta reported similar results in a study of the W3C HTML WG mailing list, revealing minimal participation from Asia, Africa, and South America and limited involvement by women \cite{Gupta2016}. Research focusing on the W3C Encrypted Media Extensions (EME) Recommendation also analyzed mailing list activity and participation in related groups, identifying comparable patterns \cite{Halpin2017, Mason2019}.

Using different indicators -- namely Recommendation editorships, participation in W3C groups, and mailing list activity -- these studies reached a consistent conclusion about the organization's composition: large companies, particularly those based in the United States and Europe, are predominant. Consequently, their interests likely shape the evolution of Web standards. However, prior research did not differentiate companies by industry sector or examine the specific topics that attract participants' attention. This paper addresses this gap by identifying W3C members' interests, as reflected in their participation in temporary groups, and by uncovering latent topics and interest clusters within the W3C. Beyond revealing the subjects driven by major Web players, our analysis also highlights areas left to other organizations, which may develop under different interests. All data and analyses are publicly available at \url{https://github.com/cewebbr/w3c_topics}.

 
\section{Data Set}
\label{sec:data}

The data used in this study is publicly available through the W3C website and API, enabling the identification of the organization to which each W3C group participant is affiliated. For each organization, we collected information on its W3C membership status and, for members, the country of their headquarters (HQ). We also employed retrieval-augmented generation (RAG, described in Sec. \ref{sec:rag}) to classify each member organization into one of the following sectors:
 \begin{itemize}
 \item \textbf{Large enterprises (LEs)}: for-profit companies with more than 250 employees.
 \item \textbf{Small and medium enterprises (SMEs)}: for-profit companies with up to 250 employees.
 \item \textbf{Non-profit organizations}: non-profit entities, including NGOs and foundations.
 \item \textbf{Government}: government institutions, including intergovernmental organizations.
 \item \textbf{Academia}: universities and research institutions.
 \item \textbf{Technical communities}: technical communities, including SDOs and open-source projects.
 \end{itemize}
These categories are primarily based on the multi-stakeholder groups defined by the Internet Governance Forum (IGF)\footnote{\url{https://www.intgovforum.org/en/about\#about-igf-faqs}} and NETmundial\footnote{\url{https://netmundial.br/2014}}, with additional subdivisions within the private sector inspired by \cite{Gamalielsson2016,Harcourt2020}. A manual verification of 60 organizations yielded an accuracy of 87\% for the RAG-generated sector classifications.

\begin{figure}[ht]
  \centering
  \includegraphics[width=0.9\columnwidth]{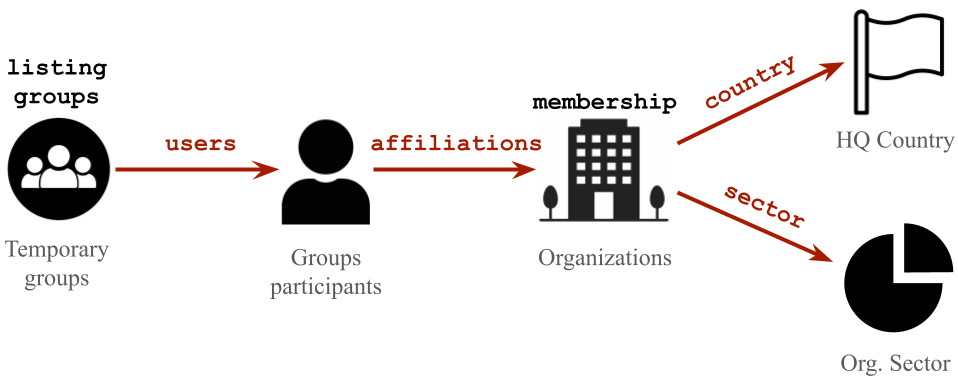}
  \caption{Graph visualization of the data relations. Datasets shown in red denote connections between entities, while those in black represent entity properties used for listing and filtering the data.}
  \Description{Graph visualization of the data relations. The link between organizations and W3C groups is established by group participants, who are affiliated to organizations.}
  \label{fig:data-structure}
\end{figure}

Table \ref{tab:sources} summarizes the data sources used in this study, along with our identification scheme and collection dates. Since we collected the data on member organizations' HQ countries several months after the other datasets, 30 organizations that had likely since left the W3C were missing from the \texttt{country} dataset. We completed these missing entries by manually consulting Wikipedia and the organizations' official websites. Figure \ref{fig:data-structure} illustrates the structure of our dataset as a graph.

\begin{table*}[ht]
  \caption{Data sources}
  \label{tab:sources}
  \begin{tabular}{lllr}
    \toprule
    \textbf{Data ID} & \textbf{Description} & \textbf{Source} & \textbf{Collection date} \\
    \midrule
    \texttt{listing} & W3C Groups listing & \url{https://api.w3.org/groups} & 2024-07-25 \\
    \texttt{groups} & Groups properties & \url{https://api.w3.org/groups/{group}} & 2024-07-25 \\
    \texttt{users} & Groups participants & \url{https://api.w3.org/groups/{group}/users} & 2024-07-25 \\
    \texttt{affiliations} & Participants' organization & \url{https://api.w3.org/users/{user_id}/affiliations} & 2024-07-25 \\
    \texttt{membership} & Org. membership status & \url{https://api.w3.org/affiliations/{org_id}} & 2024-07-26 \\
    \texttt{country} & Org. HQ country & \url{https://www.w3.org/membership/list} & 2025-04-15 \\
    \texttt{sector} & Organization sector & \textit{Our own annotation using RAG} & 2025-07-15 \\
    \bottomrule
  \end{tabular}
\end{table*}

At the time of data collection, 195 temporary groups existed: 141 CGs, 43 WGs, 9 IGs, and 2 BGs. Among them, five CGs had no W3C members. The \texttt{listing} dataset also included references to task forces -- teams assigned to specific objectives and subordinate to other W3C groups -- which were excluded from our analysis since they are not independent entities.

Group participants in the \texttt{affiliations} dataset may have up to two affiliations, one of which is always ``W3C Invited Expert.'' We standardized the data by removing this affiliation when it appeared alongside another, affecting 135 individuals from 131 organizations. This choice reflects that the ``Invited Expert'' designation is not an organization but a status that allows individuals to participate in W3C groups independently. After this standardization, 147 participants remained solely affiliated as ``W3C Invited Experts.'' Considering both W3C and this pseudo-affiliation, there were 303 member organizations with at least one representative in temporary W3C groups at the time of data collection. Approximately 21\% of W3C member organizations were not participating in any group.

\section{Methodology}
\label{sec:methodology}

\subsection{Sector annotation with RAG}
\label{sec:rag}

The sectors of W3C member organizations were annotated using retrieval-augmented generation (RAG), which combines a large language model (LLM) with external information retrieved from a vector store. We employed OpenAI's \texttt{gpt-4.1-nano} as the LLM and BGE-M3 from the Beijing Academy of Artificial Intelligence (BAAI) as the retriever embedding model. In addition to the instructions, the LLM input included the following information for each W3C member: organization name, HQ country, website and ``About'' page URLs, description extracted from the W3C website (if available), and relevant excerpts from the organization's website landing and ``About'' pages that could indicate its sector. An example of the prompt used is available in our GitHub repository.\footnote{\url{https://github.com/cewebbr/w3c_topics/blob/main/data/examples/example_rag-annotation_template.md}}

The website content was segmented into sections based on the HTML structure, with each chunk limited to 512 tokens. Excerpts were selected by the retriever using a combination of queries. First, up to two results were collected for each of the following queries: ``This is a company,'' ``This is an academic institution,'' ``This is a nonprofit organization,'' ``This is a government body,'' and ``This is a technical or standards organization.'' The retrieved results were then reranked, and the four with the highest cosine similarity scores were selected. Finally, up to two additional results from the query ``About this organization'' were added, eliminating duplicates.

\subsection{Topic modelling}
\label{sec:topic-method}

Our primary method for identifying the main topics of interest among W3C members is Latent Dirichlet Allocation (LDA), a widely used technique in Natural Language Processing (NLP) for uncovering latent themes within a collection of documents \cite{Kuo2023}. In NLP, LDA models word occurrences in each document as a stochastic process: a topic is first sampled from the document's probability distribution over $k$ topics, and then a word is sampled from the word distribution associated with that topic. By requiring all documents in the corpus to share the same set of $k$ latent word distributions, LDA infers the $k$ word distributions that best explain the observed data. The interpretation of each topic is then constructed subjectively based on its associated word distribution.

In our application of LDA, organizations play the role of documents, while the participation of an organization's representative in a given W3C group is analogous to a word occurrence (with each specific ``word'' representing a W3C group). Accordingly, we model each organization's participation in W3C as a stochastic process in which representatives are first allocated to a topic of interest and then to a W3C group, following probability distributions over groups associated with each topic. This process functions as a form of lossy information compression, summarizing each organization's interests as a mixture of $k$ topics rather than as participation across numerous groups with varying proportions. As a byproduct, the model identifies the W3C groups most commonly attended by organizations sharing a particular interest. Given the model's assumptions and the predefined number of topics $k$, the inferred latent topics -- each represented by its distribution over W3C groups -- constitute the set that best explains how W3C member organizations allocate their representatives.

Two technical aspects of our analysis warrant explanation. First, in LDA -- and in statistics more broadly -- features with low counts carry high statistical uncertainty and tend to introduce more noise than information. In our context, participation in a small W3C group may reflect an individual or organization's particular interest rather than a general or significant pattern among W3C members. Therefore, we exclude groups with a small number of participants, which are also, by definition, less representative of the most salient W3C topics.

Second, LDA does not inherently determine the number of latent topics $k$. To select this parameter, we used cross-validation (CV) by splitting the organizations into training and test sets, inferring topics from the training set, and evaluating how well their combinations represented the organizations in the test set. Specifically, we selected $k = k_{\mathrm{best}}$ that minimizes the perplexity $PP$ on the test set. Perplexity is defined as the inverse of the geometric mean of the representative allocation probabilities under the LDA model:

\begin{equation}
PP = \left[ \prod_{i=1}^N P(x_i|\bm{\Phi}, \bm{\Theta}) \right]^{-\frac{1}{N}},
\label{eq:perplexity}
\end{equation}
where $x_i$ denotes a representative allocation by an organization to a given W3C group, $N$ is the total number of participants across W3C groups, $\bm{\Phi}$ represents the distributions of groups for each topic, and $\bm{\Theta}$ represents the distributions of topics for each organization.

The choice of $k$ is influenced by the minimum group size discussed above. Including low-count groups can introduce noise and cause the topics inferred from the training set to poorly generalize to the overall dataset (i.e., overfitting), potentially resulting in the trivial case where the best estimate is the same for every organization ($k_{\mathrm{best}}=1$). Therefore, setting an appropriate minimum group size helps ensure that $k_{\mathrm{best}}>1$, meaning that the inferred topics capture multiple distinct and prevalent interests among W3C members.

\subsection{Interest similarity}
\label{sec:sim-method}

To assess whether organizations from the same sector or country share similar interests, we applied Latent Semantic Analysis (LSA), another technique commonly used in NLP \cite{Kuo2023}. In LSA, the document-term matrix (in our case, the ``organization-group'' matrix, which quantifies the number of representatives each organization allocates to different W3C groups) is normalized and approximated by a low-rank decomposition, where each document (or organization) is represented as a vector in a latent space of dimension $q$.

The normalization we applied corresponds to the $\ell$-1 \emph{term frequency-inverse document frequency} (tf-idf) scheme. First, the number of representatives each organization allocates to a group is divided by the organization's total number of representatives (tf), thereby removing the effect of organization size. This yields the fraction of representatives assigned to each group, providing a more accurate proxy for organizational interests. These fractions are then weighted by the inverse document frequency, $\mathrm{idf}(m_g)$, defined as the inverse of the fraction of organizations participating in each group, using a logarithmic scale:

\begin{equation}
  \mathrm{idf}(m_g) \equiv \log\left(\frac{M}{1 + m_g}\right) + 1,
\end{equation}
\noindent
 where $M=303$ is the total number of member organizations and $m_g$ is the number of organizations participating in group $g$. This weighting downplays groups attended by nearly all organizations, as they contribute little to distinguishing specific interests.

The low-rank approximation is obtained through truncated Singular Value Decomposition (SVD). The normalized organization-group matrix $X$ is approximated by $X'$, computed as the product of matrices $U$, $\Sigma$, and $V$, with dimensions $M \times q$, $q \times q$, and $G \times q$, respectively, where $G=190$ is the number of W3C temporary groups with at least one member organization represented:

\begin{equation}
X' = U \Sigma V^\intercal.  
\end{equation}
\noindent
The parameter $q$ determines the fraction of the total variance that the approximation captures. After decomposition, $U \Sigma$ represents each organization as a vector in a $q$-dimensional latent space. The diagonal elements of $\Sigma$, known as singular values, quantify the contribution of each dimension to explaining the variance in the data.

As a linear algebra-based method, LSA produces vectors that are less interpretable as topics than those from LDA but are more suitable for quantifying similarity. In this analysis, similarities between organizations were computed using the cosine distance between their latent vectors.

\section{Analysis}
\label{sec:analysis}

\subsection{Overall demographics}
\label{sec:demographics}

We began by assessing the level of participation of member organizations in W3C groups by counting the number of representatives each organization assigned to W3C temporary groups. This measure provides insight into which organizations may have greater advantage in nudging the development of Web standards. Although the W3C operates on a consensus-based decision-making process, in which each organization has one vote, having representatives in multiple groups can enhance an organization's ability to coordinate efforts across domains. Moreover, a larger number of representatives in a single group may strengthen an organization's influence during consensus building -- both by creating the perception of broader community support and by increasing its ability to articulate and defend positions. Nonetheless, we acknowledge that other aspects, such as technical authorship, specification editorship, and group chairing, may also significantly impact standardization outcomes.

\begin{figure}[ht]
  \centering
  \includegraphics[width=1.0\columnwidth]{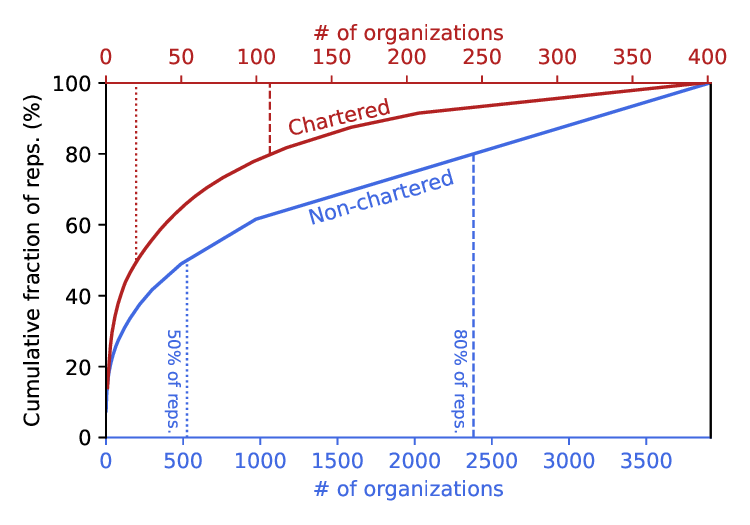}
  \caption{Cumulative fraction of organization representatives participating in chartered (red) and non-chartered (blue) groups by number of organizations, ordered by decreasing number of representatives. For chartered groups, the number of organizations is referenced on the top horizontal axis; for non-chartered groups, it is referenced on the bottom horizontal axis.}  
  \Description{Bar plot showing the number of representatives per organization. Google is the largest organization in terms of representatives, having 693. It is followed by Microsoft, with 234, and W3C invited experts, Apple, Mozilla and Intel.}
  \label{fig:reps-cum-fracs}
\end{figure}

Fig. \ref{fig:reps-cum-fracs} shows the cumulative distribution of the number of representatives per organization for chartered groups (restricted to W3C members) and non-chartered groups (open to all organizations). The vertical dotted and dashed lines indicate the number of organizations needed to account for 50\% and 80\% of all representatives, respectively. The concentration of participation in a few organizations is notably higher in chartered groups than in non-chartered ones, with Gini indices of 0.68 and 0.44, respectively. In chartered groups, half of all representatives belong to just 20 of the 397 organizations (5\%). By resampling 397 organizations from the non-chartered group data, we estimated that the probability of observing a Gini index as high as that of the chartered groups by chance is only 1.1\%, suggesting that the stronger concentration in chartered groups likely has sociological causes. A plausible hypothesis is that the W3C membership fee tends to select larger organizations or those more interested in influencing W3C standards.

\begin{figure}[ht]
  \centering
  \includegraphics[width=1.0\columnwidth]{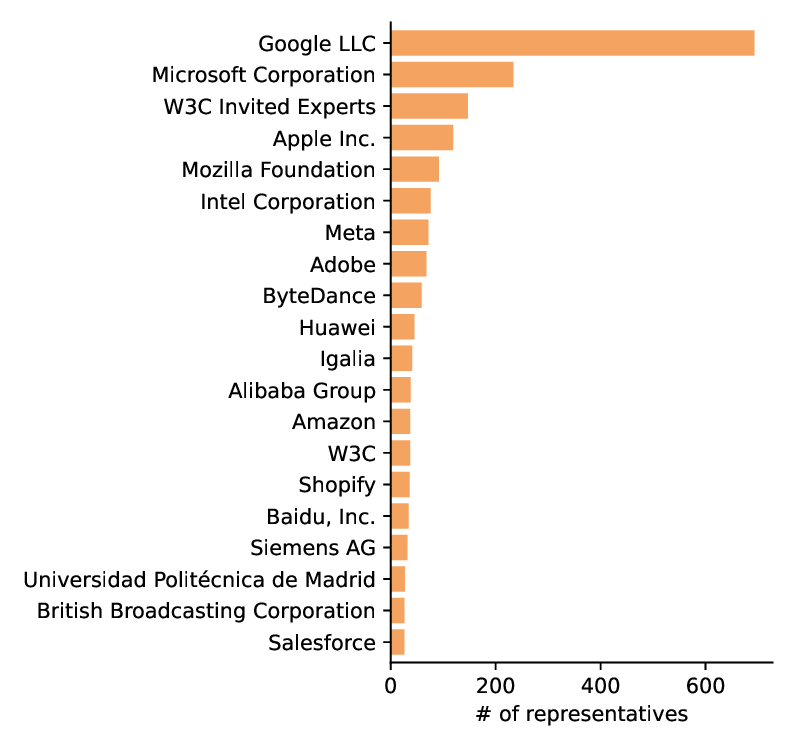}
  \caption{Number of representatives from the 20 largest organizations participating in W3C groups.}
  \Description{Bar plot shows a decreasing powerlaw-like trend. Google leads as the organization wih most representatives, with 3 times more than Microsoft, the second in the rank. It is followed by W3C Invited Experts, Apple, Mozilla, Intel, Meta, Adobe, ByteDance, Huawei, Igalia, Alibaba Group, Amazon, W3C, Shopify, Baidu, Siemens, Universidad Politécnica de Madrid, BBC, Salesforce. }
  \label{fig:reps-per-org}
\end{figure}

Fig. \ref{fig:reps-per-org} presents the number of W3C group participants affiliated with the 20 largest organizations in terms of number of representatives, all of which are W3C members. Google leads by a wide margin (693 representatives), followed by Microsoft. Notably, apart from W3C invited experts, the top four organizations are major browser vendors. 

\begin{figure}[ht]
  \centering
  \includegraphics[width=1.0\columnwidth]{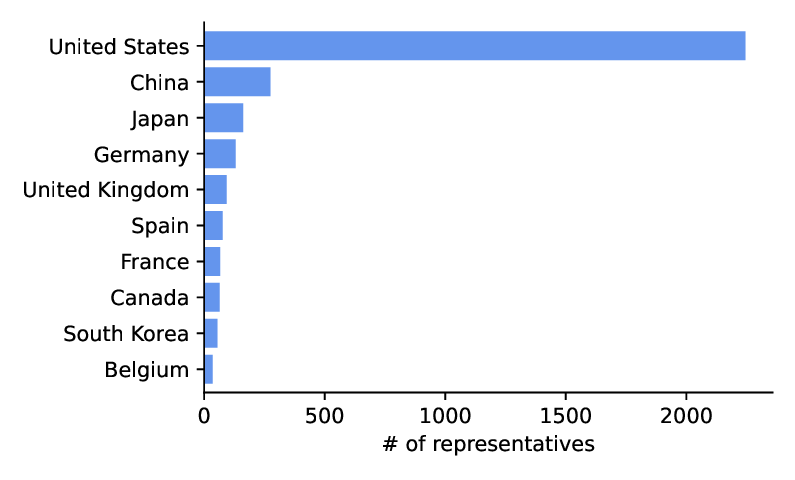}
  \caption{Number of W3C member representatives by country of HQ, showing the ten countries with the largest representations.}
  \Description{Bar plot showing a highly skewed powerlaw distribution, with United States leading with more than 8 times the next country with the most representatives: China. The latter is followed by Japan, Germany, UK, Spain, France, Canada, South Korea and Belgium.}
  \label{fig:reps-per-country}
\end{figure}

We aggregated the number of W3C member representatives by the country where each organization's HQ is located. Fig. \ref{fig:reps-per-country} presents the ten countries with the most representatives, revealing the dominant presence of organizations based in the United States. The U.S. accounts for 2,245 people, or 65.7\% of all W3C member representatives. This predominance reflects both the large number of U.S.-based W3C member organizations and the substantial size of some of these organizations. Among the top five countries, the United States, China, Japan, and the United Kingdom also host the owners of the world's most visited websites as of October 2023 \cite{Xavier2024}. In contrast, Germany is well represented among W3C members but does not appear among the 116 most visited domains, while Russia hosts several highly visited domains but has no W3C members.\footnote{Yandex participates in non-chartered groups but is not a W3C member.} 

\begin{figure}[ht]
  \centering
  \includegraphics[width=1.0\columnwidth]{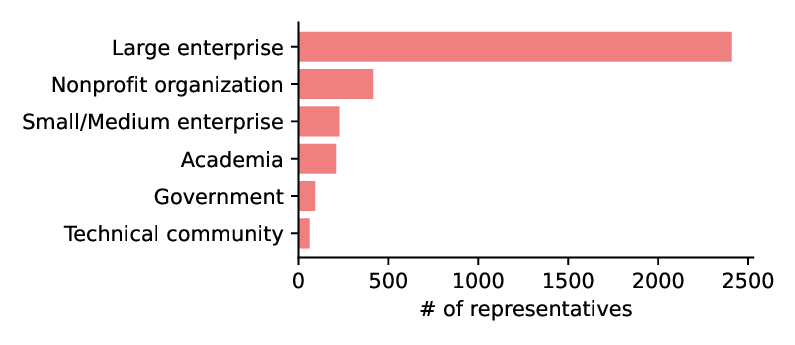}
  \caption{Number of W3C member representatives by sector.}
  \Description{Large enterprises is the sector with the most representatives, with 5 times more than Nonprofit organizations, in second place. This is followed by Small/Medium enterprises, Academia, Government and Technical Community.}
  \label{fig:reps-per-sector}
\end{figure}

Fig. \ref{fig:reps-per-sector} shows the number of W3C member representatives aggregated by sector. Representatives from LEs are by far the most common, reflecting both the high number of W3C member organizations in this sector and their tendency to be represented by larger teams. In contrast, SMEs have, on average, the fewest representatives per organization. However, their aggregate number, shown in Fig. \ref{fig:reps-per-sector}, surpasses that of Academia and Government due to the larger number of organizations in this sector.

\subsection{Topic modelling}

\begin{figure}[ht]
  \centering
  \includegraphics[width=1.0\columnwidth]{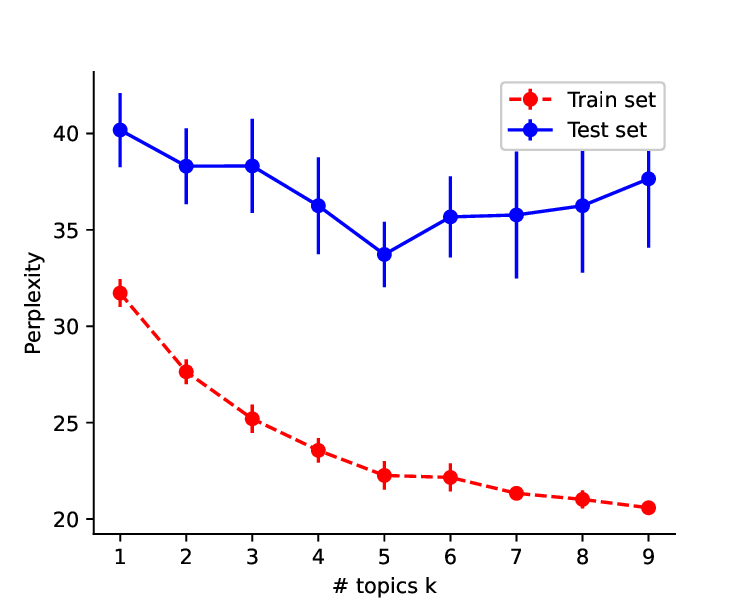}
  \caption{Perplexity averaged over CV samples used for training (red, dashed line) and testing (blue, solid line) as a function of the number of topics $k$. Error bars represent the standard deviation of the mean across CV samples.}  
  \Description{Perplexity curve for the training set goes down from k = 1 to 9, while for test sets it shows a minimum at k = 5.}
  \label{fig:perplexity}
\end{figure}

As explained in Sec. \ref{sec:topic-method}, the choice of the minimum group size cutoff $g_{\mathrm{min}}$ influences how $k$ affects the generalization capacity of the LDA model: including small W3C groups increases noise and renders the analysis uninformative. To select an appropriate $g_{\mathrm{min}}$ between 0 and 200 participants, we evaluated which value yielded the most significant decrease from $PP(k=1)$ to $PP(k=k_{\mathrm{best}})$ on the test set, indicating sharper identification of representative topics. We found that $g_{\mathrm{min}} = 55$ -- which includes the 38 largest W3C groups (about 20\% of the total) -- produced the $PP(k)$ relationship shown in Fig. \ref{fig:perplexity}.

The exact shape of the curves in Fig. \ref{fig:perplexity} depends on the pseudorandom seeds used for CV and LDA model fitting. We found that $k_{\mathrm{best}}$ typically ranged between 4 and 6, and we selected $k_{\mathrm{best}} = 5$.

\subsubsection{Topics' group composition}
\label{sec:topic-groups}

\begin{figure}[ht]
  \centering
  \includegraphics[width=1.0\columnwidth]{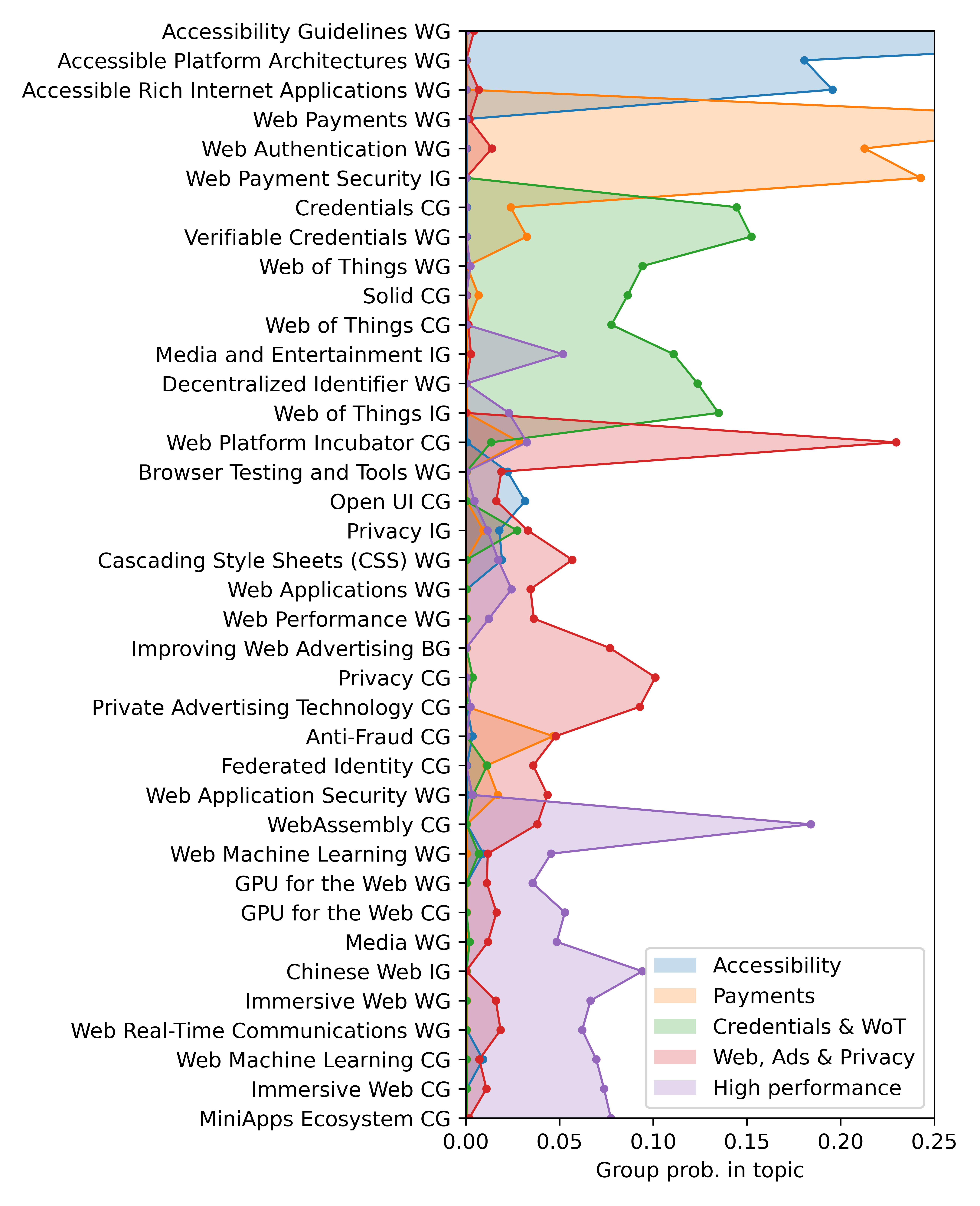}
  \caption{Probability of sampling a W3C group among the 38 largest groups for each of the five discovered topics. Each topic is represented by a different color. Probabilities above 25\% are omitted for clarity.}
  \Description{The probabilities of the groups given a topic are highly concentrated on 3 to 15 groups, and the topics have almost no overlap.}
  \label{fig:topic-group}
\end{figure}

Fig. \ref{fig:topic-group} shows the topic group distributions $\Phi$. The topics are largely disjoint, with minimal overlap along groups, indicating that each W3C group generally focuses on a specific area of interest. Notable exceptions include the Media and Entertainment IG and the Anti-Fraud CG. An interesting case involves groups explicitly related to accessibility (i.e., those mentioning it in their names). One might expect these to be transversal, contributing to multiple topics, but they appear in only one, suggesting that interest in accessibility is largely independent of other topics.

Interpreting the meaning of the topics is inherently subjective and relies on identifying connections among the most probable groups within each topic. The first two topics in Fig. \ref{fig:topic-group} (top to bottom), shown in blue and orange, consist of a small number of closely related groups, making them straightforward to characterize. One corresponds to accessibility, with 87\% of its group probability concentrated in the accessibility-focused groups. The other pertains to Web payments, with 89\% of its distribution accounted for by the Web Payments WG, Web Payment Security IG, Web Authentication WG, Anti-Fraud CG, and Verifiable Credentials (VC) WG.

The last topic, shown in purple, appears to relate to low-level hardware control (WebAssembly CG and GPU for the Web CG and WG) aimed at enabling high-performance Web applications (Immersive Web CG and WG, Web Machine Learning (ML) CG and WG, Web Real-Time Communications WG, Media and Entertainment IG, and Media WG).

The third topic combines groups related to VC (Verifiable Credentials WG, Credentials CG, Decentralized Identifier WG, and Solid CG) and the Web of Things (WoT: Web of Things IG, WG, and CG), which together account for about 81\% of the group probabilities. It also includes the Media and Entertainment IG. The overlap between VC and WoT likely arises because several research institutions and universities have representatives participating in both areas (e.g., Fundación CTIC, INRIA, ETRI, The Open University, WU - Vienna University, and FZI). Conceptually, VCs can be used to identify and authenticate things on a network \cite{Sporny2025}. The connection between the Media and Entertainment IG and WoT is likely driven by several Japanese companies -- such as MEDIA DO, ACCESS, Toshiba, NHK, and NTT -- that work with media across alternative internet-enabled electronic devices like e-books and televisions.

Finally, the fourth topic concerns core Web technologies (Web Platform Incubator CG (WICG), Cascading Style Sheets (CSS) WG, Web Applications WG, and Web Application Security WG), privacy (Privacy CG and IG, and Private Advertising Technology WG), and the Web's central business model: advertising (Private Advertising Technology CG, Improving Web Advertising BG). The combination of advertising and privacy within a single topic reflects the dependence of personalized advertising on user data, a characteristic that is prevalent on the modern Web. The existence of the Private Advertising Technology WG, whose mission ``is to specify web features and APIs that support advertising while acting in the interests of users, in particular providing strong privacy assurances using predominantly technical means''\footnote{\url{http://www.w3.org/groups/wg/pat}}, further underscores this relationship.

The relevance of advertising and privacy -- and their interconnection -- is also evident in WICG, the most prominent group within this topic. WICG is a large and diverse W3C group where new Web platform features can be proposed. Examination of its most active incubations shows that many are related to advertising, privacy, and user monitoring (e.g., turtledove, attribution-reporting-api, shared-storage, and fenced-frame).\footnote{WICG website: \url{http://wicg.io}, archived on 2025-09-17 at \url{http://archive.ph/mGjMN}.} A common feature of these proposals is their support for user privacy while maintaining the ability to monitor user behavior. Similarly, several specifications published by the Web Performance WG facilitate or focus on tracking user activity in the browser (e.g., Navigation Timing, Page Visibility, and Beacon).

To assess topic popularity, we analyzed both the effective number of member organizations interested in each topic and the effective number of representatives assigned to each topic. The former was computed by aggregating, for each topic, the organizations' probabilities of assigning a representative to that topic. The latter was obtained using the same approach, but weighting each organization's interest by the number of its representatives participating in the 38 largest groups. 

\begin{figure}[ht]
  \centering
  \includegraphics[width=1.0\columnwidth]{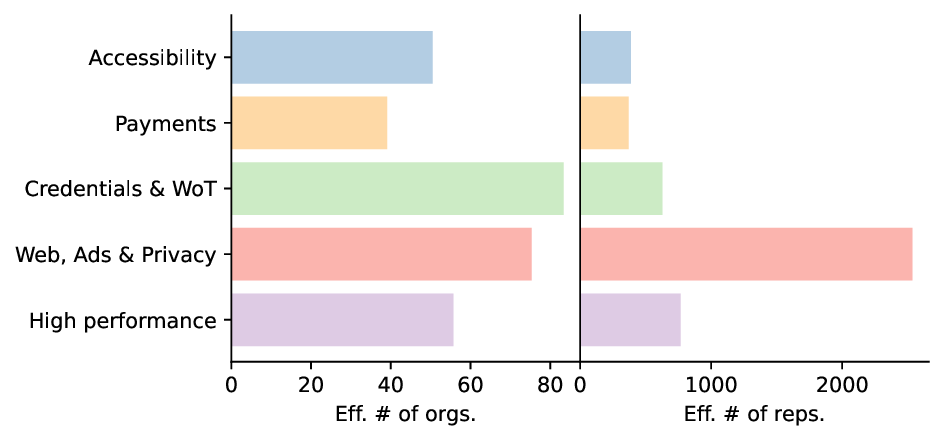}
  \caption{Effective number of organizations (left) and effective number of representatives (right) per topic.}
  \Description{The probabilities of the groups given a topic are highly concentrated on 3 to 15 groups, and the topics have almost no overlap.}
  \label{fig:topic-popularity}
\end{figure}

Fig. \ref{fig:topic-popularity} shows that topic popularity, measured by the number of organizations interested in each topic, does not vary substantially. Credentials \& WoT is the most popular topic by this metric, followed by Web, Ads \& Privacy. However, the latter far surpasses the others in terms of the number of representatives, suggesting that Credentials \& WoT attracts more interest from smaller organizations. In contrast, larger organizations tend to focus on Web, Ads \& Privacy. The slightly higher number of representatives in High Performance also indicates that it tends to be a focus of larger organizations.

\subsubsection{Organizations' topics of interest}
\label{sec:org-interests}

\begin{figure}[ht]
  \centering
  \includegraphics[width=1.0\columnwidth]{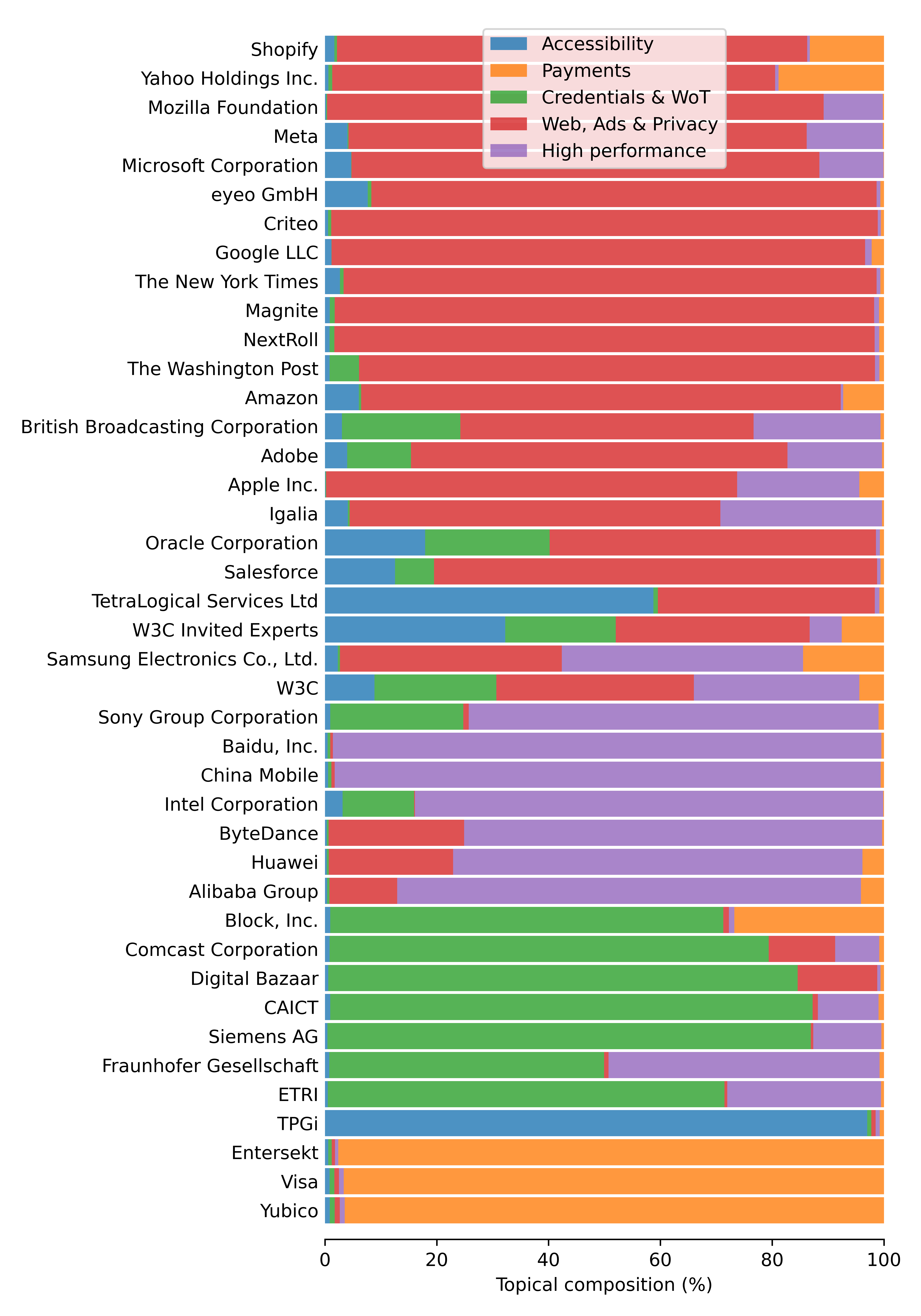}
  \caption{Topical mixture of the 41 W3C member organizations with at least 20 representatives in the 38 largest W3C temporary groups.}
  \Description{Most organizations allocate more than half of its representatives to a single topic. There are 19 organizations mostly dedicated to Web, Ads and Privacy, 7 to High performance, 7 to Credentials and Web of Things, one to Accessibility and 3 to Payments.}
  \label{fig:org-topics}
\end{figure}

LDA enables us to describe how organizations allocate their representatives -- which we use as a proxy for their interests -- as mixtures of topics. Fig. \ref{fig:org-topics} shows these mixtures for the organizations with the largest number of participants in the selected groups. Most organizations allocate more than 50\% of their representatives to a single topic; this applies to 91\% of the 266 organizations with at least one representative in the 38 largest groups. Moreover, 65\% of these organizations allocate 70\% or more of their representatives to a single topic. This suggests that most W3C member organizations primarily focus on a single topic of interest. Two exceptions serve as a coherence check for our method: W3C itself and its invited experts, whose role is to support members' work and thus participate in several temporary groups. The pattern observed in Fig. \ref{fig:topic-popularity} is also evident in Fig. \ref{fig:org-topics}: among the largest organizations, Web, Ads \& Privacy is the most prevalent topic, while Payments and Accessibility are less common.

Additional patterns emerge in Fig. \ref{fig:org-topics}. First, all major browser vendors -- Google, Microsoft, Apple, and Mozilla -- focus on the Web, Ads \& Privacy topic, with some secondary interest in High Performance. This likely reflects their interest in core browser technologies. Companies whose business models depend on advertising also concentrate on this topic, including Yahoo!, Meta, eyeo GmbH, Criteo, Google, The New York Times, Magnite, NextRoll, and The Washington Post.

Second, organizations focusing on High Performance typically include companies involved in hardware production (Sony, Intel, and Huawei), multimedia and streaming services (Sony, Baidu, ByteDance, and Alibaba Group), and telecommunications (China Mobile and Huawei). The link between high-performance technologies and these industries is reasonable. Notably, many of these organizations are Chinese -- Alibaba Group, Baidu, ByteDance, China Mobile, and Huawei.\footnote{Lenovo, another Chinese company interested in the High Performance topic, participates in the W3C but is not shown in Fig. \ref{fig:org-topics}.} However, this pattern may partly result from the assignment of the Chinese Web IG to the High Performance topic. This relationship is explored further in Sec. \ref{sec:country}.

\subsection{Similarity by sector and country}

As described in Sec. \ref{sec:sim-method}, we used LSA to measure the similarity between organizations' interests. To select the latent space dimension $q$, we created a synthetic dataset with no correlations between organizations by independently shuffling the representative counts in the organization-group matrix for each organization. By applying LSA to both the synthetic and original datasets, we identified the maximum $q$ for which the original data retained higher variance than the synthetic data, indicating how many latent components carry meaningful information about organizational similarities. We found this number to be $q = 22$.

To assess whether organizations from the same collection (sector or country) share common interests, we computed the average pairwise distance within each collection and compared it to the average pairwise distances of 10,000 randomly selected same-size sets of organizations. A significantly smaller within-collection distance compared to the random sets indicates that organizations in that collection exhibit similar interests.

To avoid the multiple comparisons problem,\footnote{\url{https://en.wikipedia.org/wiki/Multiple_comparisons_problem}} we tested only the six sectors defined in Sec. \ref{sec:data} and the three countries with the largest numbers of representatives (see Fig. \ref{fig:reps-per-country}), for a total of nine statistical tests. Assuming a significance level $\alpha_1 = 5\%$ for a single test and applying the Bonferroni correction, we set the adjusted significance level to $\alpha = \alpha_1 / n_{\mathrm{tests}} = 0.56\%$.

\subsubsection{Similarity by sector}
\label{sec:sector}

\begin{figure}[ht]
  \centering
  \includegraphics[width=1.0\columnwidth]{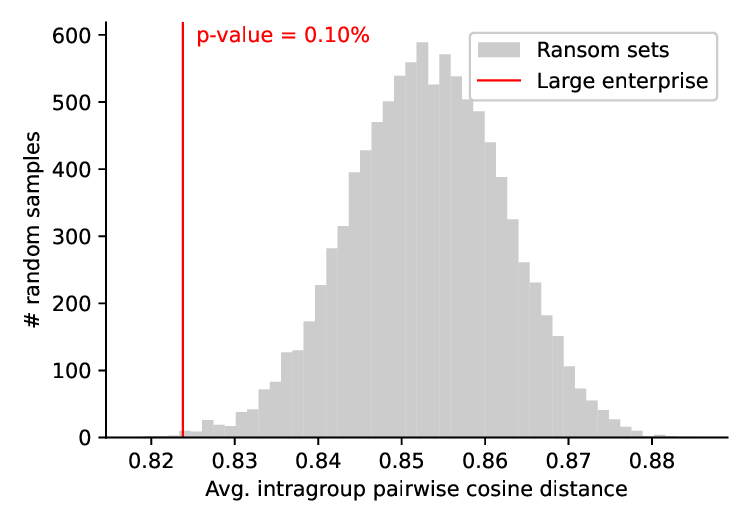}
  \caption{Histogram of the average intradistance for randomly selected sets of 127 organizations, compared with the average intradistance among the 127 large enterprise W3C members participating in temporary groups (red vertical line).}
  \Description{The histogram presents a bell shape centered near 0.85 Its left tail barely touches the red line at approximately 0.823. The p-value is 0.1\%}.
  \label{fig:LE-sim-hist}
\end{figure}

The only sector that exhibited a significantly low average intradistance was the LE sector (see Fig. \ref{fig:LE-sim-hist}), indicating that it is the only sector with confirmed shared interests. The Academia sector reached a $p$-value of 0.57\%, which is close to $\alpha$ but not below it.

To infer the common interests of the LE sector, we calculated the difference between the LE average approximated group participation and the global average, obtaining the following vector in the group space: 

\begin{equation}
  \Delta_i^{(g)} = \frac{1}{|g|}\sum_{j \in g} \frac{X'_{ji}}{|X'|_j} - \frac{1}{M}\sum_{j} \frac{X'_{ji}}{|X'|_j},
  \label{eq:avg-diff}
\end{equation}
where
\begin{equation}
  |X'|_j \equiv \sqrt{\sum_l {X'_{jl}}^2}.
\end{equation}
In Eq. \ref{eq:avg-diff}, $g$ denotes a set of organizations (in this case, LEs), and $\Delta_i^{(g)}$ represents how much W3C group $i$ is more prominent among organizations in $g$ compared to the overall average. Among LEs, the Improving Web Advertising IG stands out as particularly popular.

\subsubsection{Similarity by country}
\label{sec:country}

\begin{figure}[ht]
  \centering
  \includegraphics[width=1.0\columnwidth]{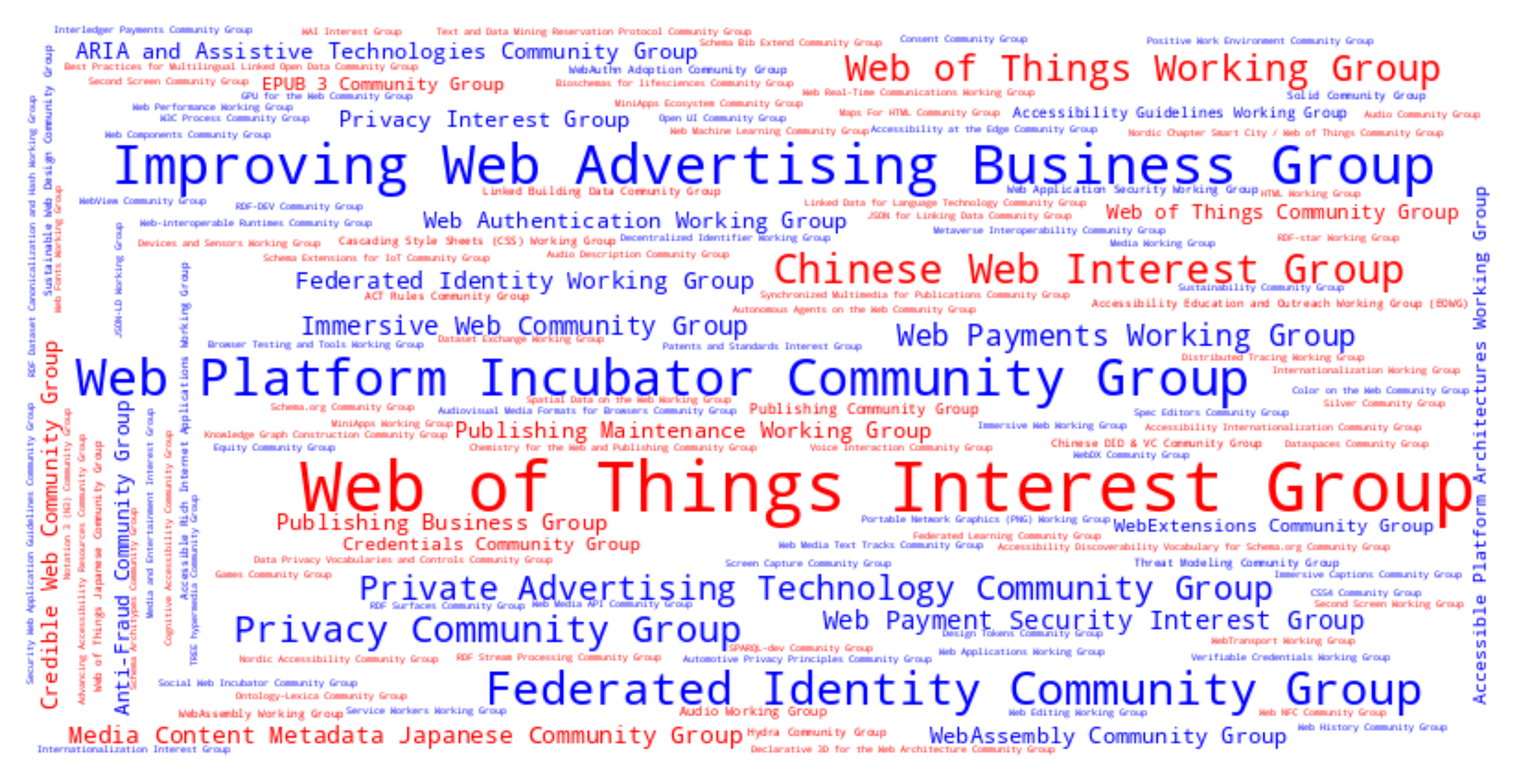}
  \caption{Word cloud of W3C temporary groups, where font size reflects $\Delta_i^{(g)}$ for organizations from the United States. Blue and red fonts indicate positive and negative $\Delta_i^{(g)}$ values, respectively.}
  \Description{In decreasing font size order, we have: Web of Things IG (in red), Improving Web Advertising BG (in blue), WICG (in blue), Federated Identity CG (in blue), Chinese Web IG (in red), Web of Things WG (in red), Privacy CG, Privacy Advertising Technology CG, and many other smaller ones.}.
  \label{fig:usa-wordcloud}
\end{figure}

We applied the same analysis to organizations from the United States, China, and Japan. In all cases, we observed significant similarity among organizations. Chinese organizations are primarily distinguished by their participation in the Chinese Web IG, while Japanese organizations focus on the Web of Things IG and the Media Content Metadata Japanese CG. In contrast, U.S. organizations show a shared interest in the Improving Web Advertising BG and the WICG, while largely disregarding the Web of Things IG. Figure \ref{fig:usa-wordcloud} represents $\Delta_i^{(USA)}$ as a word cloud.

\section{Discussion}
\label{sec:discussion}

The W3C is a highly diverse consortium of organizations, comprising participants from multiple sectors and countries working on a wide range of topics. As of July 2024, over 3,500 organizations (about 300 of which were W3C members) participated in nearly 200 temporary groups discussing subjects such as Scalable Vector Graphics, Civic Technology, and the Nordic Web of Data.

Despite this diversity, organizational representation is highly concentrated among a few entities, sectors, and countries. In chartered groups, half of all representatives are affiliated with just 20 organizations. Google stands out as the most represented organization, with 693 participants (7.8\% of the total), three times more than Microsoft, which ranks second (see Fig. \ref{fig:reps-per-org}). Regarding sector classification defined in Sec. \ref{sec:data}, large enterprises account for 69\% of W3C member representatives. This predominance of LEs and major browser vendors aligns with previous studies based on different indicators and time periods \cite{Gamalielsson2016,Harcourt2020}. In terms of HQ location, 66\% of participants represent organizations based in the United States (consistent with \cite{Gamalielsson2016}), followed by 8\% from China (see Fig. \ref{fig:reps-per-country}). This distribution mirrors the characteristics of the most visited websites as of 2023 \cite{Xavier2024}.

The inequality in representation might suggest that W3C offers little space for actors beyond major players, but our analysis indicates otherwise. W3C activities appear to be divided into largely disjoint topics, with organizations typically focusing on a single area (see Secs. \ref{sec:topic-groups} and \ref{sec:org-interests}). In other words, W3C functions more as a patchwork of independent projects than as a single, unified enterprise. In such an environment, there are always groups and topics with demographics that differ from W3C's overall composition, providing opportunities for underrepresented stakeholders to make an impact. Notable examples include the Spatial Data on the Web WG, the Dataset Exchange WG, and the Entity Reconciliation Community Group. Nevertheless, it is important to emphasize that the capacity to influence Web standards may also depend on an organization's roles, access to resources, and other factors.

By analyzing the 38 W3C groups with the largest number of member representatives, we identified the five most prominent topics of 2024 as follows:
\begin{itemize}
\item \emph{Web, Ads \& Privacy}: combining browser development and core Web technologies with personalized advertising based on user data and privacy concerns;
\item \emph{High performance}: encompassing computationally and network-intensive activities such as gaming, streaming, and running ML models that require low-level control of client hardware;
\item \emph{Credentials \& Web of Things}: a loose intersection between enabling internet connectivity for alternative devices and ensuring the identification and authentication of entities and claims on the Web or by machines;
\item \emph{Accessibility}: developing Web technologies and services that improve usability for people with various disabilities; and
\item \emph{Payments}: making Web-based payments more seamless, secure, and frictionless.
\end{itemize}

Surprisingly, we did not find significant similarity in participation by governmental institutions. This is also the case for SMEs, indicating their business is more diverse than that of LEs. For the latter, we find that they are significantly more focused on the first topic, demonstrating that advertising remains the leading business model on the Web. Interestingly, participation in ad-related groups is often accompanied by participation in privacy groups, reflecting the nature of the ad industry on the Web, based on the collection of user data. The goal of this seemingly unexpected coalition appears to be improving user privacy while preserving user data collection. 

Regarding HQ location, we observed that organizations from the United States show a preference for the first topic, which includes core Web development. This preference aligns with the fact that all four major browser vendors -- Google, Microsoft, Apple, and Mozilla -- are based in the United States. The predominance of U.S.-based companies in the advertising industry also persists beyond major players like Meta and Google. In contrast, Japanese organizations are more interested in the Web of Things, possibly reflecting their deployment of internet-connected devices, such as TVs and e-readers.

It is important to note that our analysis reflects a snapshot of W3C participation patterns and that the composition of active groups changes rapidly. With this in mind, we speculate -- based on our findings -- that the personalized advertising industry will remain strong, potentially incorporating enhanced user privacy measures; the Web platform will continue to improve in performance, enabling immersive experiences and on-device ML inference, among other capabilities; we may see expanded and automated identification and verification processes through digital media; online shopping is likely to become more seamless and effortless; and accessibility requirements and recommendations may advance further. Since W3C's work primarily consists of parallel efforts, numerous other standards and recommendations are expected to emerge, driving Web technologies in new directions as long as they do not conflict with the general trends identified here.
\balance

\bibliographystyle{ACM-Reference-Format}
\bibliography{main}

\balance

\end{document}